\documentclass[Physsubmission, Phys]{SciPost}

\hypersetup{
    colorlinks,
    linkcolor={red!50!black},
    citecolor={blue!50!black},
    urlcolor={blue!80!black}
}

\usepackage[bitstream-charter]{mathdesign}
\DeclareSymbolFont{usualmathcal}{OMS}{cmsy}{m}{n}
\DeclareSymbolFontAlphabet{\mathcal}{usualmathcal}

\begin{document}

\begin{center}{\Large \textbf{
NNLO Photon Fragmentation within Antenna Subtraction\\
}}\end{center}

\begin{center}
Thomas Gehrmann
and
Robin Sch\"urmann\textsuperscript{$\star$}
\end{center}

\begin{center}
Physik-Institut, Universit\"at Z\"urich, Winterthurerstrasse 190, CH-8057 Z\"urich, Switzerland \\
* robins@physik.uzh.ch
\end{center}

\begin{center}
\today
\end{center}


\definecolor{palegray}{gray}{0.95}
\begin{center}
\colorbox{palegray}{
  \begin{tabular}{rr}
  \begin{minipage}{0.1\textwidth}
    \includegraphics[width=35mm]{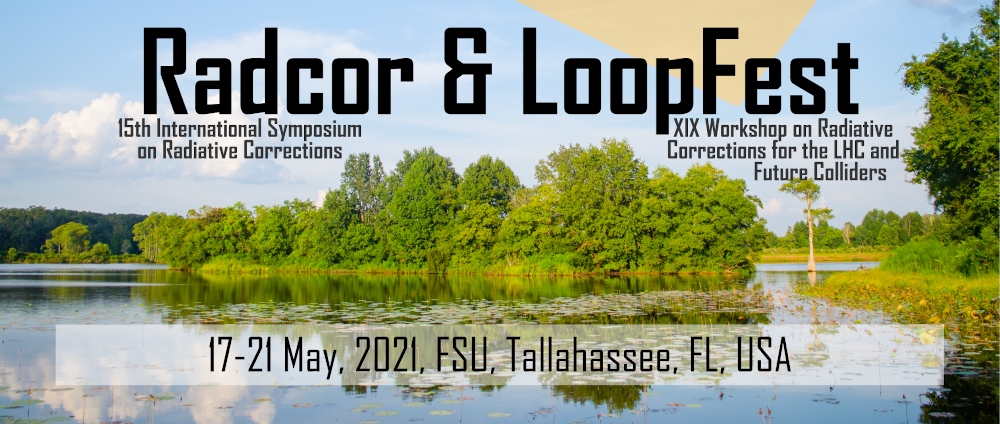}
  \end{minipage}
  &
  \begin{minipage}{0.85\textwidth}
    \begin{center}
    {\it 15th International Symposium on Radiative Corrections: \\Applications of Quantum Field Theory to Phenomenology,}\\
    {\it FSU, Tallahasse, FL, USA, 17-21 May 2021} \\
    \doi{10.21468/SciPostPhysProc.?}\\
    \end{center}
  \end{minipage}
\end{tabular}
}
\end{center}

\section*{Abstract}
{\bf
We report on our recent progress towards including the photon fragmentation contribution in next-to-next-to-leading order (NNLO) QCD predictions for photon production cross sections. This extension to previous NNLO calculations requires the identification of the photon in singular parton-photon collinear limits. We discuss how these limits can be subtracted within antenna subtraction using fragmentation antenna functions and we outline their integration.
}

\vspace{10pt}
\noindent\rule{\textwidth}{1pt}
\tableofcontents\thispagestyle{fancy}
\noindent\rule{\textwidth}{1pt}
\vspace{10pt}

\newpage

\section{Introduction}
\label{sec:intro}

The production of a photon ($\gamma$) in hadronic collisions can proceed through different mechanisms. Besides direct photons produced in the hard underlying scattering process, photons can also be produced in jet events where the jet radiates a photon in the process of hadronisation. The latter is called fragmentation contribution and is described by non-perturbative parton-to-photon fragmentation functions $D_{p \to \gamma}$~\cite{Koller:1978kq , Laermann_1982}. To reduce the contribution from fragmentation and eliminate the large background of secondary photons from hadronic decays, a photon isolation is imposed in the experimental analysis. The fixed-cone isolation limits the hadronic energy inside a fixed cone around the photon and is used in all experimental measurements to date. As it allows a limited amount of hadronic energy in the photon direction, measurements using this isolation also contain a fragmentation contribution (besides the contribution from direct photons). Alternative isolation prescriptions use a dynamical cone~\cite{Frixione:1998jh} in which the allowed hadronic energy is decreasing towards the center of the cone. These idealised isolation prescriptions fully eliminate the fragmentation contribution. \\
The calculation of isolated photon production and photon-plus-jet production cross sections at hadron colliders using a fixed-cone isolation procedure and including both, the direct and the fragmentation contribution has been achieved to next-to-leading order (NLO) QCD accuracy~\cite{AURENCHE1988661 , PhysRevD.42.61, AURENCHE199334, PhysRevD.48.3136 , PhysRevLett.73.388 , Catani_2002, Aurenche:2006vj}.
Available next-to-next-to-leading order (NNLO) QCD predictions~\cite{Campbell_2017 , Campbell_2017a , Chen_2020} for this process suppress the fragmentation contribution using an idealised isolation. Consequently, they rely on an empirical tuning of isolation parameters to mimic the experimental isolation. \\
To overcome this drawback of available NNLO QCD predictions, a calculation at this level of accuracy with a fixed-cone isolation is needed.
 Predictions with a fixed-cone isolation have to handle additional photon-parton collinear singularities which have to be extracted using a subtraction procedure, which was accomplished up to now only for $e^+e^-$ collisions~\cite{Gehrmann_De_Ridder_1998}. In contrast to singular limits in QCD,
  in which no kinematic information on individual partons has to be retained, in photonic singular limits the information on the photon momentum must not be lost. We describe how these photonic limits can be subtracted in the antenna subtraction formalism~\cite{GehrmannDeRidder:2005cm} with a new class of fragmentation antenna functions and outline their integration which remains differential in the final-state photon momentum fraction.

\section{NLO Photon Fragmentation in Antenna Subtraction}
\label{sec:NLOfrag}
To accommodate photon fragmentation in calculations based on the antenna subtraction formalism,
 we need to introduce a new type of fragmentation antenna functions. Their unintegrated forms are identical to already 
 known antenna functions, but they come with novel types of phase space factorisations, leading to different integrated antennae. 
 At NLO, only the simple-collinear quark-photon final-state singularity needs to be accounted for (and the photon is required to be 
 observed, thus preventing it from becoming soft), which implies that 
 $A_3^0(k_q,k_\gamma,k_{\bar{q}})$ in its final-final crossing and initial-final crossing with a quark in the initial state are sufficient. 
In the following we will discuss the subtraction of the quark-photon collinear limit in the initial-final configuration. \\
Using the notation established in~\cite{Currie:2013vh}, 
the real subtraction term for the photon becoming collinear to a final-state quark in the initial-final configuration reads
\begin{equation}
\begin{aligned}
{\rm d}\hat{\sigma}^S_{q(\gamma)}&= \mathcal{N}^R\sum_{{\rm perm.}}{\rm d} \Phi_{n+1}(\dots,k_q,k_\gamma,\dots;p_{q},p_2)\frac{1}{S_{n+1}}\\ &\times Q_q^2 \, A_3^0(k_q,k^{\,{\rm id.}}_\gamma,\check{p}_{\check{q}})M_n^0(\dots,k_{(q\gamma)},,\dots;\bar{p}_{q},p_2) \, J_m^{(n)}(\{\tilde{k}\}_n;z)\, ,
\end{aligned}
\label{eq:sigSNLO}
\end{equation}   
where $Q_q$ is the charge of the final-state quark and $z$  is the photon momentum fraction in the mapped momentum $k_{(q\gamma)}$. It is $z=z_3(\check{p}_{\check{q}}, k_{\gamma}^{{\rm id.}},k_q)$ with the definition of the NLO momentum fraction 
\begin{equation}
z_3\left(\check{k}_a , k_b^{{\rm id.}}, k_c\right) = \frac{s_{ab}}{s_{ab} + s_{ac}} \, .
\label{eq:defz3}
\end{equation}
The jet function  $J_m^{(n)}$ applies the jet algorithm as well as any cuts on the photon. Consequently, it retains an explicit functional 
dependence on $z$. 
The initial-state quark acts as a reference parton, indicated by the check-mark assigned to its momentum, $\check{p}_{\check{q}}$. In the initial-final configuration the reference parton is always the initial-state parton. Since the photon is identified within the cluster $k_{(q \gamma)}$, it carries the superscript '{\rm id}'. \\
To integrate the subtraction term we have to rewrite the three-body phase space appearing in the initial-final phase-space factorisation~\cite{Daleo:2006xa}  to make the integration over $z$ explicit, 
\begin{align}
  {\rm d} \Phi_{n+1}(\dots,k_q,k_\gamma,\dots;p_{q},p_2) &= {\rm d} \Phi_n(\dots, k_{(q \gamma)},\dots;\bar{p}_{q},p_2) \frac{{\rm d} x}{x} \frac{Q^2}{2\pi}{\rm d} \Phi_2   \delta\left( z -  \frac{s_{ \check{q} \gamma}}{s_{ \check{q} \gamma}+s_{\check{q}q}} \right) {\rm d} z \, ,
\end{align}
where $q^2 =  (p_{q} - k_{\gamma} - k_{q})^2=-Q^2$ and ${\rm d} \Phi_2 = {\rm d} \Phi_2(q,p_q;k_{\gamma},k_q)$.
We can integrate the initial-final real subtraction term over the unresolved phase space while staying differential in $z$. We obtain
\begin{equation}
\begin{aligned}
  \int {\rm d} \Phi_2(q,p_{q};k_{\gamma},k_q) \frac{Q^2}{2\pi} {\rm d} \hat{\sigma}^{S}_{q(\gamma)}   &= \mathcal{N}^V \frac{1}{S_n}\sum_{{\rm perm.}}{\rm d}\Phi_{n}(\dots,k_{(q \gamma)},\dots;\bar{p}_{q},p_2)\\ 
  &\times Q_q^2\mathcal{A}_{3}^{0,{\rm id.} \gamma}\left(x,z\right)M_n^0(\dots,k_{(q \gamma)},\dots;\bar{p}_{q},p_2)J_m^{(n)}(\{\tilde{k}\}_n;z)\,,
\end{aligned}
\end{equation}
with the $z$-dependent integrated fragmentation antenna function
\begin{equation}
\begin{aligned}
\mathcal{A}_{3}^{0,{\rm id.} \gamma}\left(x,z\right) &= \frac{1}{C(\epsilon)}\int {\rm d} \Phi_2(q,p_{q};k_{\gamma},k_q) A_3^0(k_q,k^{\,{\rm id.}}_\gamma,\check{p}_{q})\delta\left( z -  \frac{s_{\check{q}\gamma}}{s_{\check{q} \gamma }+s_{\check{q} q}} \right)  \\
 &= \frac{Q^2}{2} \frac{e^{\gamma_E \epsilon}}{\Gamma(1-\epsilon)} \left(Q^2\right)^{-\epsilon} \mathcal{J}(x,z) \, A_3^0(k_q,k^{\,{\rm id.}}_\gamma,\check{p}_{q}) \\
    &= \left(Q^2\right)^{-\epsilon} \left( - \frac{1}{2\epsilon} \delta(1-x) P^{(0)}_{\gamma q}(z) +\frac{1}{2} \delta(1-x) \Big(z + P^{(0)}_{\gamma q}(z) \log\big((1-z)z\big) \, \Big)   \right. \\
    &\hspace{55pt}+ \left.   \frac{1}{2} \mathcal{D}_0(x) P^{(0)}_{\gamma q}(z) - \frac{x+1}{2z} +1  \right) + \mathcal{O}(\epsilon) \, 
\end{aligned}
\label{eq:A30fragint}
\end{equation}
and the normalisation factor $C(\epsilon)=(4 \pi e^{-\gamma_E})^{\epsilon}/(8\pi^2)$. The Jacobian factor 
\begin{equation}
\mathcal{J}(x,z) = (1-x)^{-\epsilon} x^{\epsilon} z^{-\epsilon} (1-z)^{-\epsilon} \, 
\label{eq:JacPhi2}
\end{equation}
originates from expressing the integration over the two-body phase space in terms of a single integration over $z$. After expressing the invariants in the antenna function in terms of $x$ and $z$, terms of the form $(1-x)^{-1-\epsilon}$ are expanded in distributions, where we use the notation
\begin{equation}
\mathcal{D}_n(x) = \left[\frac{\log^n (1-x) }{1-x} \right]_+ \, \, , \, n \in \mathbb{N}_0 \, .
\end{equation}
As can be seen from eq.\,\eqref{eq:A30fragint}, the quark-photon collinear singularity is manifest in an $1/\epsilon$-pole at the integrated level. It is cancelled by the mass factorisation contribution from the quark-to-photon fragmentation function $D_{q\to \gamma}$, which reads
\begin{equation}
\begin{aligned}
{\rm d} \hat{\sigma}^{{\rm MF}}_{q} =& \frac{\alpha}{2\pi} {\rm d} \hat{\sigma}^B_q \otimes \boldsymbol{\Gamma}_{q\to\gamma}^{(0)}(z,\mu_A^2)\\
=& \frac{1}{2}\mathcal{N}^V \int_0^1 {\rm d} z
\sum_{{\rm perm.}} {\rm d}\Phi_n(\{k\}_n;p_q,p_2)\frac{1}{S_n}M_n^0(\dots,k_q,\dots) \, Q_q^2 \, \mu_A^{-2\epsilon} \, \Gamma_{\gamma q}^{(0)}(z) \, J_m^{(n)}(\{k\}_n;z)\,,\label{eq:FSMFantenna}
\end{aligned}
\end{equation}
where $\mu_A$ denotes the fragmentation scale  and $\Gamma^{(0)}_{\gamma q} = - (1/\epsilon) P^{(0)}_{\gamma q}$. The Born cross section at hand is
\begin{align}
{\rm d} \hat{\sigma}^B_q &= \mathcal{N}^{LO}_{{\rm jet}}\sum_{\rm perm.} {\rm d} \Phi_n(\{k\}_n;p_q,p_2)\frac{1}{S_n}M_n^0(\dots,k_q,\dots)J_m^{(n)}(\{k\}_n;z)\,.
\end{align}
In here, the jet function depends on $z$ because the quark momentum $k_q$ denotes a quark-photon cluster containing 
a photon with momentum fraction $z$. 
The normalisation factors are related through $\mathcal{N}^V=C(\epsilon)\mathcal{N}^R=2 \, C(\epsilon)(4\pi\alpha)\mathcal{N}^{LO}_\mathrm{jet}$. In eq.~(\ref{eq:FSMFantenna}) the factor 1/2 originates from different normalisation conventions of photonic and jet matrix elements. The full initial-final virtual subtraction term is then given by        
\begin{equation}
\begin{aligned}
{\rm d} \hat{\sigma}^T_{q(\gamma)} &= - \mathcal{N}^V \int_{0}^1{\rm d} z \int_0^1\frac{{\rm d}x}{x}\frac{1}{S_n}\sum_{\rm perm.}{\rm d}\Phi_{n}(\dots,k_{(q \gamma)},\dots;\bar{p}_{q},p_2)\\
&\times Q_q^2\boldsymbol{J}_{2}^{(1), {\rm id.} \gamma}(k_{(q \gamma)},\bar{p}_{q};x,z)M_n^0(\dots,k_{(q \gamma)},\dots;\bar{p}_{q},p_2)J_m^{(n)}(\{\tilde{k}\}_n;z)\,.
\end{aligned}
\label{eq:sigTIIfrag}
\end{equation}
Combination with the mass factorisation term $\Gamma^{(0)}_{\gamma q}\left(z\right)$ of the quark-to-photon fragmentation function 
yields an $\epsilon$-finite integrated fragmentation dipole: 
\begin{align}
\boldsymbol{J}_{2}^{(1), {\rm id.}\gamma}(k_{(q \gamma)},\bar{p}_{q};x,z) &= \mathcal{A}_{3}^{0,{\rm id.} \gamma}\left(x,z\right) - \frac{1}{2} \mu_A^{-2\epsilon} \, \Gamma^{(0)}_{\gamma q}\left(z\right)\delta(1-x)\,.
\end{align}

\section{Ingredients at NNLO}
\label{sec:NNLOfrag}
At NNLO, double unresolved photonic limits of double-real matrix elements as well as single unresolved photonic limits of real-virtual matrix elements have to be subtracted. The former limits correspond to triple collinear $q \parallel g \parallel \gamma$  and $q \parallel \gamma \parallel \bar{q}$ configurations. They can be subtracted using $\tilde{A}^0_4(\bar{q},g,\gamma,q)$ and $\tilde{E}^0_4(q',q,\gamma,\bar{q})$ antenna functions respectively. Their unintegrated versions have the same form as the already known antenna functions~\cite{GehrmannDeRidder:2005cm}. We use these antenna functions in the initial-final configuration, which allows to use the 
initial-state momentum as reference direction in the definition of the 
collinear momentum fraction. The double-real subtraction term for the $q \parallel g \parallel \gamma$ limit reads
\begin{equation}
\begin{aligned}
\text{d} \hat{\sigma}^{S,b_1}_{q(\gamma)} =  \mathcal{N}^{RR} &\sum_{{\rm perms}} \text{d} \Phi_{n+2}( ...,k_q   ,  k_g ,  k_{\gamma} , ... ; p_q , p_2) \frac{1}{S_{n+2}} \\
&\times \tilde{A}^0_4(\check{p}_q,k_g, k_{\gamma}^{{\rm id.}}, k_q) \, Q_q^2 \, M^0_{n+2}( ... \, , k_{(q\gamma g)} , \, ...) \, \emph{J}^{(n)}_{m} ( \{ \tilde{k} \}_n ; z) \, ,
\end{aligned}
\label{eq:sigSb1}
\end{equation}
where the momentum fraction is given by $z=z_4(\check{p}_{\check{q}}, k_{\gamma}^{{\rm id.}},k_g,k_j)$ with the definition of the NNLO momentum fraction
\begin{equation}
z_4\left(\check{k}_a,k_b^{{\rm id.}},k_c,k_d\right) = \frac{s_{ab}}{s_{ab} + s_{ac} +s_{ad}} \, .
\label{eq:defz4}
\end{equation}
$z_4$ faithfully reproduces the momentum fraction of the photon in all double unresolved limits. The $\tilde{A}^0_4$ antenna still carries a single unresolved photon limit which has to be subtracted to guarantee a cancellation of the singularities of the double-real matrix element (for details see~\cite{Currie:2013vh}). To remove this singularity an additional subtraction term is introduced which takes the form
\begin{equation}
\begin{aligned}
\text{d} \hat{\sigma}^{S,b_2}_{q(\gamma)} &=  -\mathcal{N}^{RR} \sum_{{\rm perms}} \text{d} \Phi_{n+2}(...,k_q   , k_g , k_{\gamma} , ... ; p_q , p_2) \frac{1}{S_{n+2}} \\
&\times A^0_3(\check{p}_q,k_{\gamma}^{{\rm id.}},  k_q) \, A^0_3(\check{\bar{p}}_{q},k_{g}, k^{{\rm id.}}_{(\gamma q)}) \, Q_q^2 \, M^0_{n+2}( ... \, , k_{((q\gamma) g)} , \, ...) \, \emph{J}^{(n)}_{m} ( \{ \tilde{\tilde{k}} \}_n ; z= u v)\, .
\end{aligned}
\label{eq:sigSb2}
\end{equation}
In the first mapping corresponding to the leftmost antenna function the photon is identified and a momentum fraction $u$ of the photon in the quark-photon cluster is reconstructed. In the second mapping the quark-photon cluster is identified and a momentum fraction $v$ of the cluster in the limit $g \parallel (\gamma q)$ is calculated. The momentum fraction which is used in the jet function $\emph{J}^{(n)}_{m}$ to reconstruct the photon momentum is then given by
\begin{equation}
z = u v = z_3\left(\check{p}_{\check{q}},k_{\gamma}^{{\rm id.}} , k_q\right) \, z_3\left(\check{\bar{p}}_{\check{q}}, k_{(\gamma q)}^{{\rm id.}},k_g\right) = z_4\left(\check{p}_{\check{q}}, k_{\gamma}^{{\rm id.}},k_g,k_q\right) \, ,
\end{equation}
where we used the definition of the NLO momentum fraction in eq.\,\eqref{eq:defz3} and the form of the initial-final mapping~\cite{Daleo:2006xa}. As the two momentum fractions in eq.\,\eqref{eq:sigSb1} and in eq.\,\eqref{eq:sigSb2} coincide, the cancellation of the single unresolved photon limit between the two terms is guaranteed. At NNLO, additional terms are needed for an overall subtraction of all unresolved limits of the matrix element. In these terms more $X^0_3$ fragmentation antenna functions appear where a photon-parton cluster is identified (as the second antenna function in eq.\,\eqref{eq:sigSb2}). All these fragmentation antenna functions have been integrated in the initial-final and final-final configuration~\cite{NNLO_frag}. \\
For the subtraction of the single collinear $q \parallel \gamma$ limit of one-loop matrix elements only a single one-loop antenna function, the $\tilde{A}^1_3(\bar{q},\gamma,q)$ antenna function~\cite{GehrmannDeRidder:2005cm}, is needed. We use it exclusively in the initial-final configuration. The corresponding subtraction term takes the same form as eq.\,\eqref{eq:sigSNLO} but with the replacement $\mathcal{N}^R \to \mathcal{N}^{RV}$ and $A^0_3 \to \tilde{A}^1_3$. Since we use the fragmentation antenna functions $\tilde{A}^0_4$, $\tilde{E}^0_4$ and $\tilde{A}^1_3$ in their unintegrated form in the subtraction terms, they have to be added back in their integrated form at the double-virtual level of the calculation. \\
The initial-final antenna functions are kinematically described by the scattering process
\begin{equation}
q(q^2)+p_i \rightarrow k_j + k_k (+ k_l) \, ,
\end{equation}
where $k_j^2 = k_k^2 = k_l^2 = p_i^2=0$ and $q^2= -Q^2 <0$. \\
We give a short outline of the integration of these classes of fragmentation antenna functions.

\subsection{Integration of the $X^0_4$ Fragmentation Antenna Functions}
The inclusive integrated initial-final antenna functions $\mathcal{X}^0_4$ are obtained by integration over the corresponding three-body phase space~\cite{Daleo:2009yj}, i.e.\
\begin{equation}
\mathcal{X}^0_{i,jkl}(x) = \frac{1}{C(\epsilon)^2} \int \text{d} \Phi_3(k_j,k_k,k_l;p_i,q) \frac{Q^2}{2 \pi} X^0_{i,jkl} \, ,
\label{eq:def_ifantenna_qcd}
\end{equation}
with $x= \frac{Q^2}{2p \cdot q}$ and the normalisation factor $C(\epsilon)=(4 \pi e^{-\gamma_E})^{\epsilon}/(8\pi^2)$.
For initial-final fragmentation antenna functions the same normalisation as in eq.\,\eqref{eq:def_ifantenna_qcd} is used but the integration remains differential in the final-state momentum fraction $z$, i.e.\
\begin{equation}
\mathcal{X}^{0, \, {\rm id.}  j}_{i,j k l}(x,z) = \frac{1}{C(\epsilon)^2} \int \text{d} \Phi_3(k_j,k_k,k_l;p_i,q) \, \delta\left(z -x \frac{(p_i+k_j)^2}{Q^2} \right) \frac{Q^2}{2\pi} X^0_{i,j k l} \, .
\label{eq:defintX40frag}
\end{equation}
The final-state momentum fraction is fixed by the additional $\delta$-distribution and it describes the fraction of energy carried by particle $j$ in the unresolved limit. In the definition of the momentum fraction the initial-state momentum $p_i$ is used as a reference momentum. The momentum fraction can be rewritten as $z= z_4(\check{p}_i,k_j^{\rm id.},k_k,k_l)$.
As discussed above,  all double-unresolved identified photon limits are contained in two fragmentation antenna functions: $\tilde{A}^0_4(\hat{q},g,\gamma^{{\rm id.}},q)$ and $\tilde{E}^0_4(\hat{q}',q,\gamma^{{\rm id.}},\bar{q})$.
To integrate these fragmentation antenna functions, we use the reduction to master integrals technique. Using unitarity relations, we can express the $\delta$-distributions in eq.\,\eqref{eq:defintX40frag} by cut propagators~\cite{Anastasiou:2002yz}. In this way we rewrite the phase-space integrals as $2\to 2$ three-loop integrals in forward scattering kinematics. \\
The reduction to master integrals is performed using the program \texttt{Reduze2}~\cite{vonManteuffel:2012np}, which uses the Laporta algorithm~\cite{Laporta:2000dsw} to solve the system of equations between the different integrals obtained by integration-by-parts techniques~\cite{Tkachov:1981wb, Chetyrkin:1981qh} and Lorentz invariance~\cite{Gehrmann:1999as}. \\
The initial-final scattering kinematics gives rise to 12 propagators from which four are cut propagators. In the reduction to master integrals we require all integrals to contain the four cut propagators in the denominator. For the integration of the two photonic $X^0_4$ fragmentation antenna functions we find nine master integrals. The master integrals are calculated using their differential equations in the two kinematic variables $x$ and $z$. The boundary conditions are fixed by integrating the solution of the differential equations over $z$ and comparing the result with the inclusive master integrals in~\cite{Daleo:2009yj}. The expressions for the integrated fragmentation antenna functions are very lengthy~\cite{NNLO_frag} so that we do not quote them here.  

\subsection{Integration of the $X^1_3$ Fragmentation Antenna Functions}

The inclusive integrated one-loop antenna functions in the initial-final configuration are defined as~\cite{Daleo:2009yj}
\begin{equation}
\mathcal{X}^1_{i,jk}(x) = \frac{1}{C(\epsilon)} \int \text{d} \Phi_2(k_j,k_k;p_i,q) \frac{Q^2}{2 \pi} X^1_{i,jk} \, ,
\label{eq:def_X31_QCD}
\end{equation}
where $X^1_{i,jk}$ is the unintegrated one-loop antenna function and $\text{d} \Phi_2$ the two-particle phase space. We define the integrated  initial-final one-loop fragmentation antenna functions in line with eq.\,\eqref{eq:def_X31_QCD} as 
\begin{equation}
\begin{aligned}
\mathcal{X}^{1, {\rm id.} j}_{i,j k}(x,z) &= \frac{1}{C(\epsilon)} \int \text{d} \Phi_2(k_j,k_k;p_i,q) \, \delta \left( z - \frac{s_{i j}}{s_{i j} + s_{i k}} \right) \frac{Q^2}{2 \pi} X^1_{i,j k} \\
&= \frac{Q^2}{2} \frac{e^{\gamma_E \epsilon}}{\Gamma(1-\epsilon)} \left(Q^2\right)^{-\epsilon} \mathcal{J}(x,z) \, X^1_{i,j k} \, .
\end{aligned}
\label{eq:def_X31_integrated_photonic}
\end{equation}
The integration takes the same form as for the $X^0_3$ fragmentation antenna functions, see eq.\,\eqref{eq:A30fragint} above.
 The Jacobian factor $\mathcal{J}$ is given in eq.\,\eqref{eq:JacPhi2}.
As can be seen from eq.\,\eqref{eq:def_X31_integrated_photonic}, no actual integration has to be performed to obtain the integrated fragmentation antenna functions $\mathcal{X}^{1, {\rm id.} j}_{i,j k}$. However, to express the integrated fragmentation antenna functions in terms of distributions in $(1-x)$ and in $z$ we first have to cast the unintegrated antenna functions in a form suitable for this expansion. Therefore, deriving the integrated initial-final one-loop fragmentation antenna functions follows the steps of the derivation of the integrated initial-initial one-loop antenna functions presented in~\cite{Gehrmann:2011wi}. In contrast to the NLO $X^0_3$ antenna functions which only contain rational terms in the invariants, the one-loop antenna functions $X^1_3$ also contain logarithms and polylogarithms in the invariants. These functions have branch cuts in the limits $x \to 1$ and $z \to 0$. Therefore, the expansion in distributions in $z=0$ and $x=1$ cannot be performed directly. We follow the strategy of~\cite{Gehrmann:2011wi} and express the one-loop antenna functions in terms of one-loop master integrals. \\
The one-loop master integrals appearing in the expressions for the one-loop antenna functions are the one-loop bubble ${\rm Bub}(s_{ij})$ and the one-loop ${\rm Box}(s_{ij},s_{ik})$ in all kinematic crossings. 
Both master integrals are well-defined in the Euclidean region, in which all invariants are smaller than 0. The master integrals have to be analytically continued from this kinematic region, to the kinematic region under consideration given by
\begin{equation}
s_{ij} < 0 \quad , \quad s_{ik} < 0 \quad , \quad s_{jk} >0 \quad , \quad s_{ijk} = - Q^2 < 0 \, .
\end{equation}
While the analytic continuation of the bubble master integrals is trivial, care has to be taken in case of the box integrals where  hypergeometric functions appear. In the analytic continuation of these box integrals the hypergeometric functions must not have a branch cut in the kinematic endpoints $x=1$ and $z=0$ so that an expansion in distributions can be performed. The $z$-integration of the resulting expressions recovers
the known real-virtual initial-final master integrals~\cite{Daleo:2009yj} and enabled us to identify an error in their numerical implementation for 
jet production in deep-inelastic scattering~\cite{Currie:2017tpe}.
 \\
The relevant one-loop integrated fragmentation antenna function for photon production is \break $\mathcal{\tilde{A}}^{1, \, {\rm id.} \gamma}_3(x,z)$. Its expression is very lengthy~\cite{NNLO_frag} such that we do not quote it here.

\subsection{Coefficient Functions for Semi-Inclusive Deep Inelastic Scattering}

The antenna functions $A^0_3$, $A^1_3$ and $\tilde{A}^1_3$ are derived from the squared matrix element of the scattering $\gamma^{\star} \to q g \bar{q}$ at tree level and at one-loop respectively and the antenna functions $A^0_4$ and $\tilde{A}^0_4$ are derived from the squared matrix element of the scattering process $\gamma^{\star} \to q g  g\bar{q}$. Therefore, the results for the integrated version of these fragmentation antenna function can be compared to the coefficient functions of semi-inclusive deep inelastic scattering. Using the notation of~\cite{Abele:2021nyo}, the expansion of the semi-inclusive coefficient functions reads
\begin{equation}
\omega^i_{f' f} = \omega_{f' f}^{i,(0)}(x,z) + \frac{\alpha_s}{\pi} \omega^{i,(1)}_{f'f}(x,z) + \left( \frac{\alpha_s}{\pi} \right)^2 \omega^{i,(2)}_{f'f}(x,z) + \mathcal{O}(\alpha_s^3) \, ,
\end{equation}
with $i=T,L$. $f'$ corresponds to the parton which fragments into the final-state hadron and $f$ corresponds to the parton in the initial state. We find that the combination of coefficient functions corresponding to the integrated fragmentation antenna functions is $\omega^T - \omega^L/(2- 2\epsilon)$. The results for the one-loop coefficient functions are well known~\cite{Altarelli:1979kv}. Comparing the results of the integrated $X^0_3$ fragmentation antenna functions to the results stated in~\cite{Anderle:2012rq}, we find
\begin{equation}
\begin{aligned}
\frac{1}{4  Q_q^2 C_F} \left( \omega^{T,(1)}_{gq} - \frac{1}{2(1-\epsilon)} \omega^{L,(1)}_{gq} \right) &= \mathcal{A}^{0,{\rm id.} \gamma}_3\left(x,z\right) - \mu_A^{-2\epsilon} \frac{1}{2} \Gamma^{(0)}_{\gamma q}(z) \, , \\
\frac{1}{4 Q_q^2 C_F} \left( \omega^{T,(1)}_{qq} - \frac{1}{2(1-\epsilon)} \omega^{L,(1)}_{qq} \right) &= \mathcal{A}^{0, {\rm id. q}}_3\left(x,z\right) - \mu_F^{-2\epsilon} \Gamma^{(1)}_{qq}(x) - \mu_A^{-2\epsilon} \Gamma^{(1)}_{qq}(z) + \mathcal{A}^1_2 \, ,
\end{aligned}
\end{equation}
where $\mathcal{A}^1_2$ is the one-loop vertex factor~\cite{GehrmannDeRidder:2005cm} in space-like kinematics. The NNLO coefficient functions $\omega_{f'f}^{(2)}$ are unknown, however, recently an approximation of the NNLO coefficient functions $\omega^{i,(2)}_{qq}$ was derived using the threshold resummation formalism~\cite{Abele:2021nyo}. \\
With the results of our integrated fragmentation antenna functions we are able to predict the subleading color contribution to the above combinations of the semi-inclusive coefficient functions $\omega^{i,(2)}_{gq}$. Expressed in terms of the integrated antenna functions and mass factorisation kernels it reads
\begin{equation}
\begin{aligned}
&\frac{1}{4 Q_q^2 C_F} \left( \omega^{T,(2)}_{gq} - \frac{1}{2(1-\epsilon)} \omega^{L,(2)}_{gq} \right) \bigg\rvert_{1/N} \\
&= \tilde{\mathcal{A}}^{0, {\rm id. \gamma}}_4(x,z) + \tilde{\mathcal{A}}^{1 \, U, {\rm id.} \gamma}_3(x,z) - \mu_A^{-2\epsilon} \frac{1}{2} \Gamma^{(0)}_{\gamma q}(z) \otimes \left( \mathcal{A}^{0, {\rm id.} q}_3(x,z) + \mathcal{A}^1_2 \right) \\
&- \mu_F^{-2\epsilon} \Gamma^{(1)}_{qq}(x) \otimes \mathcal{A}^{0, {\rm id.} \gamma}_3(x,z) + \frac{1}{2} \mu_A^{-2\epsilon} \Gamma^{(0)}_{\gamma q}(z) \otimes \left( \mu_F^{-2\epsilon} \Gamma^{(1)}_{qq}(x) + \mu_A^{-2\epsilon} \Gamma^{(1)}_{qq}(z) \right) \\
&- \frac{1}{2} \left(\mu_A^2\right)^{-2\epsilon} \Gamma^{(1)}_{\gamma q}(z) \, , 
\end{aligned}
\end{equation}
where the notation $...|_{1/N}$ means that only terms proportional to $1/N$ are considered. The expressions for the initial-state mass factorisation kernel $\Gamma^{(1)}_{qq}$  can be found in~\cite{Currie:2013vh} 
and the one-loop photon fragmentation mass factorisation kernel $\Gamma^{(1)}_{\gamma q}$ is given by~\cite{Gehrmann_De_Ridder_1998}:
\begin{equation}
\Gamma^{(1)}_{\gamma q}(z) = \frac{1}{4\epsilon^2} \left(P_{qq}^{(0)} \otimes P_{\gamma q}^{(0)}\right) (z) -\frac{1}{4\epsilon} P^{(1)}_{\gamma q}(z) \, .
\end{equation}

\section{Conclusion}
In this talk, we have described the extension of the antenna subtraction method to account for photon fragmentation processes up to NNLO,
deriving all newly required phase space factorisations and integrated antenna functions~\cite{NNLO_frag}. New results on NNLO 
semi-inclusive DIS coefficient functions were obtained as a by-product. 
Our extension of the antenna subtraction method 
will allow to compute NNLO corrections to processes with final-state photons at hadron colliders for the realistic
 fixed-cone isolation prescriptions used in the experimental measurements, thereby improving upon previous results for the 
idealised dynamical-cone isolation.

\section*{Acknowledgements}
We would like to thank Marius H\"ofer and Alexander Huss for interesting discussions on the work presented here. 
This work was supported by the Swiss National Science Foundation (SNF) under contract 200020-175595.

\bibliography{bib_NNLO_frag.bib}

\nolinenumbers

\end{document}